\documentclass[preprint,amsmath,amssymb]{revtex4}
\usepackage{graphicx}
\usepackage{dcolumn}
\usepackage{bm}
\usepackage{hyperref}
\usepackage{amsmath}
\usepackage{amssymb}

\begin{document}
\title{Manipulation of optical memory bits in
atomic vapors and Bose-Einstein condensates}

\author{ I.P. Vadeiko}
\address{School of Physics and Astronomy, University of St
Andrews, North Haugh, St Andrews, KY16 9SS, Scotland}

\author{A.V. Rybin }
\address{Department of Physics, University of Jyv\"askyl\"a PO
Box 35, FIN-40351 Jyv{\"a}skyl{\"a}, Finland}

\author{ A. R. Bishop}
\affiliation{Theoretical Division and Center for Nonlinear Studies,
Los Alamos National Laboratory, Los Alamos, New Mexico 87545, USA}

\begin{abstract}
We provide an exact analytic  description of decelerating, stopping
and re-accelerating optical solitons in atomic media. By virtue of
this solution we describe  in detail how spatially localized optical
memory bits can be written down, read  and moved along the atomic
medium in a prescribed manner. Dynamical control over the solitons
is realized via a background laser field whose intensity controls
the velocity of the slow light in  a similar way as in the linear
theory of electromagnetically induced transparency (EIT). We solve
the nonlinear model   when the controlling field and the solitons
interact in an inseparable   nonlinear superposition process. This
allows us to access results beyond the limits of the linear theory
of EIT.
\end{abstract}
 \maketitle

Today, significant theoretical and experimental efforts are made to
better understand the physics of light propagation in atomic vapors
and Bose-Einstein condensates whose interaction with light is well
described by the nonlinear $\Lambda$-model. This process  is
accompanied by striking effects such as electromagnetically induced
transparency~\cite{Lukin:2001}, which permits the propagation of
light through an otherwise opaque medium; slow-light phenomena when
the pulse is slowed down to velocities of a few meters per
second~\cite{Hau:1999}, and optical information storage when the
pulse is completely stopped inside the dielectric~\cite{Bajcsy:2003,
Dutton:2004}. Even though the linear approach to describing these
effects is developed in detail, it has reached its validity limits,
because modern experiments require more adequate nonlinear
descriptions. However, the theory of the nonlinear regime is still
very incomplete.

We investigate the problem of control over optical memory bits
created by the slow-light solitons. Our theoretical model
corresponds to a gas of alkali atoms, whose interaction with light
is approximated by the structure of levels of the $\Lambda$-type
(see Ref.~\cite{Lukin:2001}). The two lower levels are the states
$|1\rangle$ and $|2\rangle$, while $|3\rangle$ is the excited state.
The medium is described by the $3\times3$ density matrix $\rho$ in
the interaction picture. In order to cancel the residual Doppler
broadening, two optical beams are chosen to be co-propagating.  The
fields are described by the Rabi-frequencies $\Omega_{a,b}$. The
field $\Omega_a$ corresponds to $\sigma^-$ polarization, while the
second $\Omega_b$ corresponds to $\sigma^+$ polarization. Within the
slowly varying amplitude and phase approximation (SVEPA),  dynamics
of the atom-field system is well described by the reduced
Maxwell-Bloch equations \cite{Rybin:2004}:
\begin{eqnarray}\label{Maxwell_Bloch}
\partial_\zeta H_I&=&i\frac {\nu_0}4
\left[{D,\rho}\right],\;D=\left(%
\begin{array}{ccc}
  1 & 0 & 0 \\
  0 & 1 & 0 \\
  0 & 0 & -1 \\
\end{array}%
\right),\nonumber\\
\partial_\tau \rho&=&i\left[{\frac\Delta2 D-
    H_I,\rho}\right].
\end{eqnarray}
Here $\zeta=(z-z_0)/c$, $\tau=t-(z-z_0)/c$, $\Delta$ is the detuning
of the resonance, and $\nu_0$ is the coupling constant. The matrix
$H_I=-\frac12 \left({\Omega_a |3\rangle\langle1|
+\Omega_b|3\rangle\langle2|}\right) +h.c.$ represents the
interaction Hamiltonian.

The system of equations Eq.(\ref{Maxwell_Bloch}) is exactly solvable
in the framework of the Inverse Scattering (IS) method
\cite{fad,Rybin:2004,Park:1998,gab}. We define the initial and
boundary conditions for the system of equations
Eq.(\ref{Maxwell_Bloch}) as follows. We consider a semi-infinite
$\zeta\ge0$ active medium with a pulse of light
 incident at the point $\zeta=0$ (initial condition). This means
 that the evolution is considered with respect to the {\it space}
 variable $\zeta$, while the boundary conditions should be specified
 with respect to the  variable $\tau$. The boundary conditions
 are defined by the asymptotic values of the density matrix $\rho$
 and the matrix $H_I$ at $\tau\to\pm\infty$.

In  our previous  work \cite{Rybin:2004} we proposed a method to
realize quantum memory bits in BEC by  slow-light optical solitons.
There we reported a slow-light soliton solution on a constant
background field $\Omega_0$, which was created by the auxiliary
laser on entrance into the medium ($\zeta=0$).  The group velocity
of the slow-light soliton depends explicitly on the field
$\Omega_0$. For the time-dependent background field some approximate
solutions based on the methods of scaled time are studied in
\cite{Grobe:1994, Bajcsy:2003, Dey:2003, ulf:2004}. In a simplifying
approximation the velocity is $v_g\approx
c\frac{\Omega_0^2}{2\nu_0}$. This expression immediately suggests
that the soliton can be stopped through switching off the
controlling laser field \cite{Rybin:2004,ryb7}. In the present paper
we provide a full analytical description of how one can perform
controlled preparation, manipulation and readout of optical memory
bits formed in atomic vapors and  Bose-Einstein condensates by
slow-light solitons.

Specifically,  we consider a gas of rubidium atoms. To make
parameters dimensionless, we measure the time in units of optical
pulse length $t_p=1\mu s$ typical for the experiments on the
slow-light phenomena in rubidium vapors \cite{Bajcsy:2003,
Kash:1999}. The retarded time $\tau$ is measured in microseconds and
the Rabi frequencies are normalized to $\mathrm{MH}z$. The spatial
coordinate is normalized to the spatial length of the pulse slowed
down in the medium, i.e. $l_p=v_g t_p\approx c \frac
{\Omega_0^2}{2\nu_0}t_p$. In experiments with rubidium the
controlling background field is of order of a few megahertz. We
choose $\Omega_0=3$. This corresponds to the group velocity of
several meters per second, depending on the density of the atoms. We
take the group velocity to be $10^{-7} c$, so the pulse spatial
length is $30 \mu m$, and $\zeta$ is normalized to $10^{-13} s$.
Then, in the dimensionless units, the coupling constant $\nu_0=\frac
{\Omega_0^2}2=4.5$.

\begin{figure}[htb]
\centerline{\includegraphics[width=70mm]{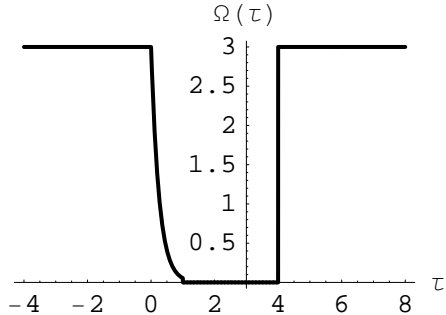}}
\caption{\label{fig1}   Time-dependence of the controlling
background field. We choose $\alpha=4$, and the delay interval is
$T-T_1=3$. The dimensionless units are defined  in the text.}
\end{figure}

We consider the following scenario for the dynamics (see
Fig.\ref{fig1}). To begin with, in a remote past we create in the
medium a slow-light soliton, and assume it is propagating on the
constant background $\Omega_0$. In the next step, we slow down the
soliton by switching off the laser source of the background field.
We assume an exponential decay of the background field with a decay
constant $\alpha$, i.e. $\Omega_0e^{-\alpha\tau}$. At a certain, say
$T_1=4/\alpha$, the field becomes negligible. Therefore, we cut off
the exponential tail and approximate it by zero. At this step the
soliton is completely stopped. The position where the soliton stops,
depends on the decay constant $\alpha$ and  on the moment when we
switch off the laser  \cite{ryb7}. The process of stopping the
soliton corresponds to writing a memory bit into the medium. The
information borne by the soliton is stored in the form of spatially
localized polarization. This formation can live a relatively long
period of time in atomic vapours or BEC. At the moment of time $T$
we restore the slow-light soliton by abruptly switching on the
laser. The whole dynamics is divided into four time intervals
$\bigcup_{i=0}^3{\cal
D}_i=(-\infty,0]{\cup}[0,T_1]{\cup}(T_1,T]{\cup}(T,\infty)$.
 The time-dependence of the intensity of the background field at
entrance into the medium is given in Fig.\ref{fig1}.

Before  the soliton enters the medium, the atoms are assumed to be
in the state $|1\rangle$. Notice that this state is a dark-state for
the controlling field, which means that the atoms do not interact
with the field created by the auxiliary laser. We specify the
initial state of the system as follows:
\begin{equation}\label{init_fields0}
    \Omega^{(0)}_a=0,\; \Omega^{(0)}_b=\Omega(\tau),\;
    |\psi_{at}\rangle=|1\rangle.
\end{equation}
This configuration corresponds to a typical  experimental setup (see
e.g. \cite{Hau:1999,Liu:2001,Bajcsy:2003}). The function
$\Omega(\tau)$ is  given in Fig.\ref{fig1}. The state
Eq.(\ref{init_fields0}) satisfies the Maxwell-Bloch equations
Eqs.(\ref{Maxwell_Bloch}). Using the methods of our previous works
\cite{Rybin:2004, ryb7}, we construct the one-soliton solution
corresponding to the background profile (see Fig.\ref{fig1}), viz.

\begin{eqnarray}\label{fields_tilde}
\tilde\Omega_a&=&\frac{(\lambda^*-\lambda)w(\tau,\lambda)
} {\sqrt{1+|w(\tau,\lambda)|^2}}\; e^{i\tilde\theta_s}\, \mathrm{sech}\tilde\phi_s,\\
\tilde\Omega_b&=&\frac{(\lambda-\lambda^*)w(\tau,\lambda) }
{1+|w(\tau,\lambda)|^2}\,e^{\tilde\phi_s}\,
\mathrm{sech}\tilde\phi_s-\Omega(\tau),\nonumber
\end{eqnarray}
with the atomic state $\tilde\rho=|\tilde\psi_{at}\rangle\langle
\tilde\psi_{at}|$, where
\begin{eqnarray}\label{atoms_tilde}
|\tilde\psi_{at}\rangle=\frac{\mathrm{Re}\lambda-\Delta-i
\mathrm{Im}\lambda\tanh\tilde\phi_s} {|\lambda-\Delta|}  |1\rangle +
\frac{\tilde\Omega_a}{2|\lambda-\Delta|w(\tau,\lambda)} |2\rangle-
\frac{\tilde\Omega_a}{2|\lambda-\Delta|} |3\rangle.\quad
\end{eqnarray}
Here,
\begin{eqnarray}\label{params_tilde}
\tilde\phi_s&=&\tilde\phi_0+
\frac{\nu_0\zeta}{2}\mathrm{Im}\frac1{\lambda-\Delta}+
\mathrm{Re}(z(\tau,\lambda))+\ln\sqrt{1+|w(\tau,\lambda)|^2},\nonumber\\
\tilde\theta_s&=&\tilde\theta_0-\frac{\nu_0\zeta}2 \mathrm{Re}
\frac1{\lambda-\Delta}+\mathrm{Im}(z(\tau,\lambda)),\nonumber
\end{eqnarray}
\noindent where $\lambda$ is an arbitrary complex parameter. The
functions $w(\tau,\lambda)$, $z(\tau,\lambda)$ are of  piecewise
form, specific to each time region ${\cal D}_i$.

\begin{table*}[htb]
  \centering
  \caption{\label{table1}Exact analytical solution}\begin{tabular}
  {|c|c|c|c|c|}\hline
 ${\cal D}$& $\Omega(\tau)$ & $w(\tau,\lambda)$ & $z(\tau,\lambda)$ & ${\cal C}$
 \\ \hline\hline

${\cal D}_0$ & $\Omega_0$ & $w_0$
   & $\frac i2 \Omega_0 w_0 \tau$
    & 0 \\ \hline

${\cal D}_1$ & $\Omega_0 e^{-\alpha\tau}$ &
    $i\frac{{{\cal C}} J_{1-\gamma}
\left(-\frac{\Omega(\tau)}{2\alpha}\right)- J_{\gamma-1}
\left(-\frac{\Omega(\tau)}{2\alpha}\right)}{{\cal C} J_{-\gamma}
\left(-\frac{\Omega(\tau)}{2\alpha}\right)+ J_{\gamma}
\left(-\frac{\Omega(\tau)}{2\alpha}\right)}$ &
$-\alpha\gamma\tau+\ln \frac{{{\cal C}} J_{-\gamma}
\left(-\frac{\Omega(\tau)}{2\alpha}\right)+ J_{\gamma}
\left(-\frac{\Omega(\tau)}{2\alpha}\right)}{{\cal C} J_{-\gamma}
(-\frac{\Omega_0}{2\alpha})+ J_{\gamma}
\left(-\frac{\Omega_0}{2\alpha}\right)}$ & $\frac{-i w_0 J_{\gamma}
(-\frac{\Omega_0}{2\alpha})+ J_{\gamma-1}
(-\frac{\Omega_0}{2\alpha})}{ J_{1-\gamma}
(-\frac{\Omega_0}{2\alpha})+ i w_0 J_{-\gamma}
(-\frac{\Omega_0}{2\alpha})}$ \\\hline

${\cal D}_2$ & $0$ & 0 & $\ln\frac{{\cal C}
\left(-\frac{\Omega_0}{4\alpha}\right)^{-\gamma}/\mathrm{\Gamma}(1-\gamma)}
{{\cal C} J_{-\gamma} (-\frac{\Omega_0}{2\alpha})+ J_{\gamma}
\left(-\frac{\Omega_0}{2\alpha}\right)}$ & ${\cal C}_2={\cal C}_1$
\\\hline

${\cal D}_3$ & $\Omega_0$ & $\frac{\Omega_0
\tan\left({\frac12\sqrt{\lambda^2+\Omega_0^2}(\tau-T)}\right)}
{\lambda\tan\left({\frac12\sqrt{\lambda^2+\Omega_0^2}(\tau-T)}\right)
-i\sqrt{\lambda^2+\Omega_0^2} }$ & $\begin{array}{c}\ln\frac{{\cal
C}\, e^{\frac{-i\left({\lambda+\sqrt{\lambda^2+\Omega_0^2}}\right)
(\tau-T)}2}+ e^{\frac{-i
\left({\lambda-\sqrt{\lambda^2+\Omega_0^2}}\right)
(\tau-T)}2}}{{\cal C}+1}\\+z_2\end{array}$
 & $\frac{\Omega_0^2+2\lambda
\left({\lambda-\sqrt{\lambda^2+\Omega_0^2}}\right)}
{\Omega_0^2}$ \\
\hline
  \end{tabular}
\end{table*}
For clarity we organize elements of the solution corresponding to
different time regions in Table~\ref{table1}. We use an auxiliary
function
$w_0={\Omega_0}/({\lambda+\sqrt{\lambda^2+{\Omega_0}\!^2}})$, the
index  of Bessel functions is defined as
$\gamma=({\alpha+i\lambda})/({2\alpha})$.
 The values ${\cal C}_i $ of the constant ${\cal C}$ for each time region ${\cal D}_i$ are specified in the
 rightmost column of the table, the moment of time $T$  is chosen as in Fig.\ref{fig1},
 i.e. $T=4/\alpha+3$. Notice that in the table $w_2=w_1(\infty,\lambda)$
 and $z_2=z_1(\infty,\lambda)$. Therefore the solution in the region ${\cal D}_2$ is
parameterized by the asymptotic values of the data for the region
${\cal D}_1$ corresponding to  the absence of cut-off of the
exponentially vanishing tail. The region ${\cal D}_2$ describes the
phase when the slow-light soliton is stopped, the fields vanish,
while the information borne by the soliton is stored in the medium
in the form of spatially localized polarization. At the moment of
time $T$ the laser is instantly turned on again. The stored
localized polarization then generates  a moving slow-light soliton.
This process is described by the solution in the region ${\cal
D}_3$. Except for the point $T_1$, the functions $w,z$ are
continuous in $\tau$. This ensures  that the physical variables such
as the wave-function and field amplitudes evolve continuously.
\begin{figure}[thb]
\centerline{\includegraphics[width=70mm]{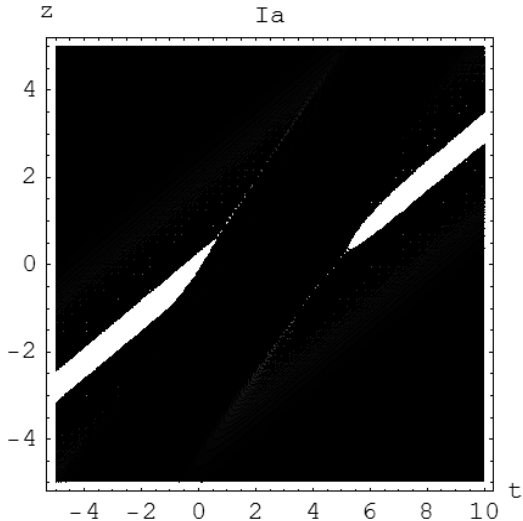}}
\caption{\label{fig2} Contour plot of the intensity of
$\tilde\Omega_a$. We choose $\lambda=-4.1i$ and zero detuning,
$\Delta=0$. The break-up area in between the two solitonic trails
manifests the creation of a standing memory bit in the medium.}
\end{figure}

\begin{figure}[thb]
\centerline{\includegraphics[width=80mm]{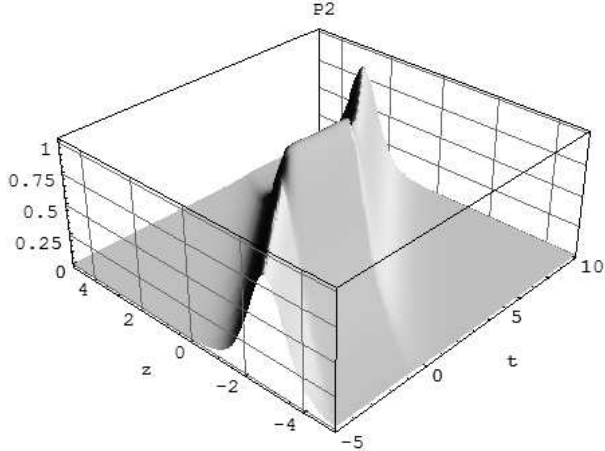}}
\caption{\label{fig3} Population of level $|2\rangle$. Here,
$\lambda=-4.1i$ and $\Delta=0$. The  time interval $1 \le t \le 4$
corresponds to a standing localized  polarization flip.}
\end{figure}

We demonstrate typical dynamics of the intensity of the field in the
channel $a$ in Fig.\ref{fig2}.    This  contour plot shows that in
the process of rapid deceleration the solitonic trail forms   end
sharply. It is interesting to note that the restored pulse, however,
has the same characteristics, i.e. the width and group velocity, as
the input signal existed in the medium before the stopping. The
dynamics of the corresponding localized polarization is presented in
Fig.\ref{fig3}. Notice  that in the presence of the soliton the
population flip from  level $|1\rangle$ to $|2\rangle$ in the center
of the peak is almost complete. A small fraction of the total
population is located in the upper level $|3\rangle$ and provides
for some atom-field interaction. In \cite{Rybin:2004} we show that
the population of  level $|3\rangle$ is proportional to
$|\Omega_0|^2$.  Therefore when the field vanishes the population of
the second level reaches  unity, while the destructive influence of
relaxation becomes negligible. This behavior of the system points to
a sound possibility for realizing  optical memory.

To conclude, with an exactly solvable example, we have demonstrated
a possibility to manipulate  memory bits by  slow-light solitons. We
have proposed a method   to prepare, control and read out optical
memory bits in atomic vapors and Bose-Einstein condensates in the
regime of strong nonlinearity.

IV acknowledges the support of the Engineering and Physical Sciences
Research Council, United Kingdom. Work at Los Alamos National
Laboratory is supported by the USDoE.


\end{document}